\documentclass{article}

\usepackage{amsmath}
\usepackage{graphicx,psfrag,epsf}
\usepackage{enumerate}
\usepackage{natbib}
\usepackage{url} 
\usepackage{bm}
\usepackage{algorithm}
\usepackage{mathptmx}
\usepackage{lmodern}
\usepackage{amssymb,amsmath,amsthm,bm} 
\usepackage{subcaption}
\usepackage{bbm}
\usepackage{multirow,multicol}
\usepackage{longtable}
\usepackage{physics}
\usepackage{rotating}
\usepackage{verbatim,booktabs}
\usepackage[mathscr]{euscript}

\usepackage{algpseudocode}

\newtheorem{prop}{Proposition}

\def\spacingset#1{\renewcommand{\baselinestretch}%
{#1}\small\normalsize} \spacingset{1}

\title{Mixture of Finite Mixtures Model for Basket Trial}
\author{Junxian Geng, Tianjian Zhou, Ruitao Lin, Guanyu Hu}

\date{}

\begin{document}

\maketitle

\begin{abstract} 
With the recent paradigm shift from cytotoxic drugs to new generation of target therapy and immuno-oncology therapy during oncology drug developments, patients with various cancer (sub)types may be eligible to participate in a basket trial if they have the same molecular target. Bayesian hierarchical modeling (BHM) are widely used in basket trial data analysis, where they adaptively borrow information among different cohorts (subtypes) rather than fully pool the data together or doing stratified analysis based on each cohort. Those approaches, however, may have the risk of over shrinkage estimation because of the invalidated exchangeable assumption. We propose a two-step procedure to find the balance between pooled and stratified analysis. In the first step, we treat it as a clustering problem by grouping cohorts into clusters that share the similar treatment effect.  In the second step, we use shrinkage estimator from BHM to estimate treatment effects for cohorts within each cluster under exchangeable assumption. For clustering part, we adapt the mixture of finite mixtures (MFM) approach to have  consistent  estimate of  the  number of  clusters. We investigate the performance of our proposed method in simulation studies and apply this method to Vemurafenib basket trial data analysis.

\end{abstract}
\textbf{Bayesian Hierarchical Modeling, Clustering Detection, Nonparametric Bayesian Method, Oncology}

\newpage
\spacingset{1.45} 
\section{Introduction}\label{sec:intro}

With the recent paradigm shift from cytotoxic drugs to new generation of target therapy and immuno-oncology therapy during oncology drug developments, clinical trials in oncology no longer solely target for a specific cancer type based on the anatomic location of the primary tumor (e.g., breast, lung or GI). Patients with various cancer (sub)types may be eligible to participate in a clinical trial if they have the same molecular target. The term basket trial is used to represent those studies where a single targeted or immuno-onocology therapy is investigated in the context of multiple diseases or disease subtypes \citep{woodcock2017master}.  

For oncology drug development, we usually start patients studies in Phase I. The phase I study is typically comprised of two parts, Phase Ia and Phase Ib. The objective in Phase Ia is to find Maximum Tolerated Dose (MTD) or Optimal Biological Dose (OBD) based mainly on safety data with the supplements of initial efficacy information, and we usually call this part dose escalation. After the optimal dose is determined, we enroll more patients to that dose level to mainly evaluate efficacy while still monitoring safety of the drug, where we name as expansion part of the study (Phase Ib). During Phase Ib, basket trial design are widely used to evaluate the efficacy of the new treatment in multiple cohorts. 

Assume single-arm settings for all cohorts in phase Ib without control arm, a binary endpoint of response (yes/no) is usually used. There are different criterion for various tumor types. For example, RECIST criteria \citep{eisenhauer2009new} is used for solid tumor and RANO criteria \citep{lin2015response} is used for brain tumor. Even with slight different definitions among different criterion, response is typically defined as a tumor shrinkage more than a certain threshold (e.g. 30$\%$ in RECIST criteria). The reasonable positive correlation with time-to-event endpoints such as Progression Free Survival (PFS) and Overall Survival (OS) along with much earlier readout, makes the binary endpoint of response widely acceptable efficacy endpoint in early phase oncology studies. Eventually the treatment effect will be reflected in response rate, which is defined as the number of patients with responses divided by the total number of patients. Thus without further discussion on the validity of the endpoint, a binary endpoint is assumed in this paper. 

When dealing with multiple cohorts in one study, there are two directions as we conduct data analysis: pooled analysis or stratified analysis. If the assumption is the treatment effect is homogeneous among all cohorts, a pooled analysis may be conducted where we get a response rate by combining the data from all cohorts. If the assumption is that different cohorts have totally heterogeneous treatment effect, separate response rate will be estimated in each cohort. In practice, however, both of these two assumptions might not be true. It's very hard to justify a drug has homogeneous effect on different tumor types, but it's also difficult to claim that totally different treatment effects are expected among those cohorts taking the same drug. In addition, fully pooled analysis might dilute the positive signal if the drug works only in some of the cohorts and fully stratified analysis might be either lack of power or costly inefficient. 

In order to mitigate the above mentioned problems, several methods were proposed to find a balance between pooled analysis and stratified analysis, where they adaptively borrow information among different cohorts rather than fully pool the data together. Many of those methods are based on Bayesian hierarchical modeling (BHM). \cite{thall2003hierarchical} and \cite{berry2013bayesian} proposed exchangeable assumptions among the parameters (logit transformation of response rate) for different cohorts, which led to shrinkage estimators for each cohort. Those exchangeable assumptions, however, can easily lead to over shrinkage to a common mean, thus introduce bias to the estimators. \cite{neuenschwander2016robust} and \cite{chu2018bayesian} further allow each cohort‐specific parameter to be exchangeable with other similar strata parameters or nonexchangeable with any of them in order to mitigate the risk of over shrinkage. These methods sometimes may suffer from too many pre-specified parameters.

In this article, we propose a two-step procedure to find the balance between pooled and stratified analysis in basket trial design. In the first step, we treat it as a clustering problem by grouping cohorts into clusters that share the similar treatment effect. In the second step, we use shrinkage estimator proposed by \cite{berry2013bayesian} to estimate treatment effects for cohorts within each cluster under exchangeable assumption. For clustering, we adapt the mixture of finite mixtures (MFM) approach \citep{miller2018mixture,geng2019probabilistic,geng2020discovery}, which admits a clustering scheme similar to the famous Chinese restaurant process (CRP) but alleviates the drawback of CRP by automatic model-based pruning of the tiny extraneous clusters leading to consistent estimate of the number of clusters. The contribution of this paper is three-fold. First, a full Bayesian framework is developed and the clustering results yield useful probabilistic interpretation. In addition, we establish consistency results for the estimation of clusters for the binary data. Thirdly, high probability of selecting the correct number of clusters and the better estimation performance than benchmark methods when heterogeneity exits between cohorts are empirically demonstrated.

The rest of the paper is organized as follows. We start with a brief review of the existing popular models for basket trial design and MFM model, as well as propose our two-step procedures in Section~\ref{sec:method}. The Bayesian methods for simultaneous inference on the number of clusters and cluster-specific estimates are discussed in Section~\ref{sec:bayesian} and the MCMC algorithm is provided in Section~\ref{sec:MCMC}. Simulation studies and comparisons with existing methods are provided in Section~\ref{sec:simu} and illustrations of our method by two case studies are presented in section \ref{sec:real_data}. Finally, we have some discussions in Section~\ref{sec:discussion}.

\section{Methodology}\label{sec:method}

\subsection{Existing models for basket design}\label{existing}
Suppose there are $N$ cohorts in the proposed basket design. In cohort $i$, there are $n_i$ patients enrolled with response rate $p_i$. If the number of responses in cohort $i$ is denoted as $r_i$, we have:
\begin{align}
& r_i \mid n_i, p_i \stackrel{\text{ind}}
\sim
\mbox{Binomial}(n_i, p_i), \quad i =
1,
\ldots, N
\end{align}
In a single arm setting in early phase oncology trials, the goal is to get an estimation of $p_i$ for each cohort $i$ and compare that estimate to the benchmark response rate threshold. If the estimated response rate is greater than the pre-specified threshold in a cohort, the sponsor will consider continuing the development of the tested drug for that indication.  

A popular Bayesian hierarchical modeling used to get estimations of $p_i$'s is proposed by \cite{berry2013bayesian}, where they use shrinkage estimators for each cohort under exchangeable assumptions and we call this approach Berry's BHM in the rest of the paper. Then model and prior can be briefly expressed as:
\begin{align}\label{eq:berry}
& \theta_i = \text{log}(\frac{p_i}{1-p_i}) - \text{log}(\frac{p_{Ti}}{1-p_{Ti}}) 
\\ \nonumber
& \theta_i \sim \mbox{N}(\mu,\sigma^2), \quad i =
1,
\ldots, N,  \\ \nonumber
& \mu \sim \mbox{N}(0,2^2), \quad \sigma^2 \sim \mbox{HN}(0,1), \\ \nonumber
& r_i \mid n_i, p_i \stackrel{\text{ind}}
\sim
\mbox{Binomial}(n_i, p_i), \quad i =
1,
\ldots, N
\end{align}
Where $p_{Ti}$ is usually set as the benchmark response rate or control rate for cohort $i$. $\theta_i$ can be understood as the logit transformed treatment effect relative to the control rate. The values of the hyperparameters in aforementioned model for $\mu$ and $\sigma^2$ are popular used and they are generally not too sensitive to different choices. 

The performance of this model, however, is highly dependent on the validity of the exchangeable assumptions as well as the choices of control rate, which sometimes is hard to justify in real studies. If the exchangeable assumptions are not valid for $\theta$'s, there is high risk of over shrinkage on the estimations of cohort-specific response rate.  

\cite{neuenschwander2016robust} proposed a method that further allow each cohort‐specific parameter to be exchangeable with other similar strata parameters or nonexchangeable with any of them, which is a data-drive approach to mitigate the risk of over shrinkage and we call this approach EXNEX in the rest of the paper. The model and prior can be briefly expressed as:
\begin{align}\label{eq:exnex}
& \theta_i = \text{log}(\frac{p_i}{1-p_i})
\\ \nonumber
\mbox{EX: with probability } \pi_i; \quad & \theta_i \sim \mbox{N}(\mu,\sigma^2), \quad i =
1,
\ldots, N,  \\ \nonumber
& \mu \sim \mbox{N}(a,b^2), \quad \sigma^2 \sim \mbox{HN}(c,d), \\ \nonumber
\mbox{NEX: with probability } 1-\pi_i; \quad & \theta_i \sim \mbox{N}(\mu_i,\sigma^2_i), \quad i =
1,
\ldots, N,  \\ \nonumber
& r_i \mid n_i, p_i \stackrel{\text{ind}}
\sim
\mbox{Binomial}(n_i, p_i), \quad i =
1,
\ldots, N
\end{align}
Where $\pi_i$ is the weight of exchangeable component for cohort $i$, which is usually set the same for all cohorts by default; the hyperparameters $a, b, c, d$ for the exchangeable part is usually from the method for setting weakly informative priors given in the on-line Appendix of \cite{neuenschwander2016robust} unless there are specific prior information; the hyperparameters $\mu_i$ and $\sigma^2_i$ for each cohort are usually set so that $\mu_i$ equals to the log-odds of a plausible guess (or benchmark response rate) for cohort $i$ and $\sigma^2_i$ makes approximately one observation. 

This model sometimes may suffer from too many pre-specified parameters or parameters that need to be tuned.

\subsection{Mixture of Finite Mixtures model for binary data}
To mitigate the possible issues mentioned above, we first cluster the cohorts into groups based on the observed response rates and then get shrinkage estimators within each group by assuming the exchangeable assumption. 

Bayesian models offer a natural solution to simultaneously estimate the number of clusters and cluster assignments. The Chinese restaurant process \citep[CRP;][]{neal2000markov} offers choices to allow for uncertainty in the number of clusters by assigning a prior distribution on the cluster assignments parameters $(z_1, z_2, \ldots, z_N)$, where $z_i$ denotes the cluster assignment for cohort $i$. In the CRP, $z_i, i=2, \ldots, N$ have the following conditional distribution  \citep[i.e., a P\'{o}lya urn scheme,][]{blackwell1973ferguson}: 
\begin{eqnarray}\label{eq:crp}
P(z_{i} \mid z_{1}, \ldots, z_{i-1})  \propto  
\begin{cases}
\abs{c}  , \quad  \text{at an existing cluster labeled}\, c\\
\alpha,  \quad \quad \quad \, \text{at a new cluster}.  
\end{cases}
\end{eqnarray}
Here $\abs{c}$ refers to the size of cluster labeled $c$, and $\alpha$ is the concentration parameter of the underlying Dirichlet process. 
The prior distribution on the cluster assignments induces a prior distribution on the sizes of the clusters in the partition. Let $\mathcal{C}_N$ denotes a partition of the set $\{1, 2, 3, \ldots,N\}$ and $t = \abs{\mathcal{C}_N}$ denote the number of blocks in the partition $\mathcal{C}_N$. 

Under \eqref{eq:crp}, one can obtain the probability of block-sizes $\bm{b}= (b_1, b_2, \ldots, b_t)$ of a partition $\mathcal{C}_N$ as:
\begin{eqnarray} \label{eq:sizeprobcrp}
p_{\mathrm{DP}}(\bm{b}) \propto  \prod_{j=1}^t  b_j^{-1}.  
\end{eqnarray}

It is clear from \eqref{eq:sizeprobcrp} that CRP tends to assign large probabilities to highly imbalanced cluster sizes in which, necessarily, some clusters will be quite small. This results in producing extraneous clusters in the posterior that leads inconsistent estimation on the \textit{number of clusters} even when the sample size goes to infinity.

\cite{miller2018mixture} proposed a modification to the CRP, which is called a mixture of finite mixtures (MFM) model, to mitigate this issue:  
\begin{eqnarray}\label{eq:MFM}
\begin{split}
k & \sim p(\cdot), \\
\pi = (\pi_1, \ldots, \pi_k) \mid k &\sim \mbox{Dir}(\gamma, \ldots, \gamma), \\
 z_i \mid k, \pi & \sim \sum_{s=1}^k  \pi_s \delta_s, \quad  i=1, \ldots, N, 
\end{split}
\end{eqnarray}
where $p(\cdot)$ is a proper probability mass function on $\{1, 2, \ldots\}$, and $\delta_s$ is a point-mass at $s$.  
The joint distribution of $(z_1, \ldots, z_N)$ under \eqref{eq:MFM} admits a P\'{o}lya urn scheme akin to the CRP:
\begin{eqnarray}\label{eq:mcrp}
P(z_{i}\mid z_{1}, \ldots, z_{i-1})  \propto  
\begin{cases}
\abs{c} + \gamma  , \quad  \text{at an existing cluster labeled}\, c\\
\frac{V_i(t+1)}{V_i(t)} \gamma,  \quad \quad \quad \, \text{at a new cluster}, 
\end{cases}
\end{eqnarray}
where the coefficients $V_i(t)$ are given by:
\begin{align*} 
\begin{split}
V_i(t) &= \sum_{k=1}^{+\infty}\dfrac{k_{(t)}}{(\gamma k)^{(i)}} p(k)
\end{split}					
\end{align*} 
where $k_{(t)}=k(k-1)...(k-t+1)$, and $(\gamma k)^{(i)} = {\gamma k}(\gamma k+1)...(\gamma k+i-1)$.
While this restaurant process bears close resemblance to the CRP, the introduction of new clusters is slowed down by a factor $V_i(\abs{\mathcal{C}_{i-1}}  +1)/ V_i(\abs{\mathcal{C}_{i-1}})$, thereby pruning the tiny extraneous clusters.  
 
An alternative way to understand the natural pruning of extraneous clusters is through the probability distribution induced on the cluster sizes. Again, let $\mathcal{C}_N$ denotes a partition with block-sizes $\bm{b} = (b_1, b_2, \ldots, b_t)$ and $t= |\mathcal{C}_N|$ under MFM. In contrast to \eqref{eq:sizeprobcrp}, the probability of the cluster sizes $(b_1, \ldots, b_t)$ under the MFM is:
\begin{eqnarray} \label{eq:sizeprob}
p_{\mathrm{MFM}}(\bm{b}) \propto  \prod_{j=1}^t  b_j^{\gamma-1}.  
\end{eqnarray} 
From \eqref{eq:sizeprobcrp} and \eqref{eq:sizeprob}, the comparison easily reveals that MFM assigns comparatively smaller probabilities to highly imbalanced cluster sizes.    

We adapt MFM to our model setting in order to group different cohorts into clusters, then the model and prior can be expressed hierarchically as: 
\begin{align}\label{eq:basket_MFM}
& k \sim q(\cdot), \text{where $q(\cdot)$ is a p.m.f on \{1,2, \ldots\} }
\nonumber \\
& P_{s} \stackrel{\text{ind}} \sim \mbox{Beta}(\alpha, \beta),  
\quad s = 1, \ldots, k,  \nonumber
\\ \nonumber
& \mbox{pr}(z_i = s \mid \bm{\pi}, k) = \pi_s, \quad s = 1, \ldots, k, \, i =
1,
\ldots, N,  \\ \nonumber
& \bm{\pi} \mid k \sim \mbox{Dirichlet}(\gamma, \ldots, \gamma),\\ 
& r_i \mid n_i,\bm{z},\bm{P}, k \stackrel{\text{ind}}
\sim
\mbox{Binomial}(n_i, p_i), \quad p_{i} = P_{z_i},\, i =
1,
\ldots, N, 
\end{align}
We assume $q(\cdot)$ is a $\mbox{Poisson}(1)$ distribution truncated to be
positive through the rest of the paper, which has been proved by
\cite{miller2018mixture} and \cite{geng2019probabilistic} to guarantee
consistency for the mixing distribution and the number of clusters. Here, we define $G = \sum_{s=1}^k \pi_s \delta_{P_s}$, where $\delta$ is the point mass measure, and $P_s$ is the collection of parameters for the binomial distribution in cluster $s$ for $s=1,\ldots,k$ and $P$ is Beta probability measure. 

To ensure the finite mixture of Binomial distributions is strongly identifiable up to label switching, we assume the sample size for each cohort satisfies $\min_{i=1,\ldots,N} n_i \geq 2k_{max} - 1$, where $k_{max}$ is an assumed finite upper bound on the true number of clusters $k_0$.

Let $k_0$, $G_0$, $P_0$ be the true number of clusters, the true mixing measure, and the corresponding probability measure, respectively. Then the following proposition establishes the posterior consistency and contraction rate for $k$ and $G$.

\begin{prop}\label{thm1}
Let $\Pi_N(\cdot \mid r_1,\ldots,r_N)$ be the posterior distribution obtained from \eqref{eq:basket_MFM} given a random sample $r_1,\ldots,r_N$. Assume Assumption 1 holds and that all the parameters are restricted to a compact space $\Theta^* \subset (0,1)$. Then we have:
\begin{align*}
\Pi_N(k = k_0 \mid r_1,\ldots,r_N) \rightarrow 1, ~~ \Pi_N (W(G,G_0)\lesssim (\log N/N)^{-1/2} \mid r_1,\ldots,r_N) \rightarrow 1, 
\end{align*}
almost surely under $P_0$ as $N \rightarrow \infty$, where $W(G, G_0)$ denotes the Wasserstein distance.
\end{prop}

\begin{proof}
In order to prove Proposition \ref{thm1}, we systematically verify the conditions (P.1)-(P.4) outlined in \cite{guha2019posterior} and \cite{yin2020analysis} for overfitted finite mixture models:
\begin{itemize}
    \item \textbf{(P.1) Strong Identifiability and Metric:} While finite mixtures of general discrete distributions are not unconditionally identifiable, Teicher (1963) and Blischke (1964) establish that a mixture of $k$ Binomial distributions is identifiable if the number of trials $n_i \geq 2k - 1$. Under Assumption 1, first-order identifiability is satisfied. Because the parameter space $\Theta^*$ is compact, the Wasserstein metric $W(G, G_0)$ bounds the Hellinger distance of the marginal distributions.
    \item \textbf{(P.2) Prior Positivity:} We assign a Beta base distribution to the component parameters $P_s$, which is strictly continuous and possesses full support on $(0,1)$. Because the true parameters lie in the interior of the bounded support $\Theta^*$, any Kullback-Leibler (KL) neighborhood around the true mixing measure $G_0$ contains strictly positive prior probability mass.
    \item \textbf{(P.3) Bounded Metric Entropy:} The parameter space for $P_s$ is a subset of a compact interval, and the simplex for the mixing weights $\bm{\pi}$ is finite-dimensional. Consequently, the covering number of the space of mixing distributions under the Wasserstein metric grows at most polynomially, satisfying the required exponentially bounded metric entropy condition.
    \item \textbf{(P.4) Prior Tail on $k$:} We utilize a truncated Poisson distribution for $q(k)$. This naturally satisfies the strict exponential tail decay condition $P(k > j) \leq C \exp(-c j \log j)$ for some constants $C, c > 0$, preventing the posterior from placing excess mass on unnecessarily large dimensions.
\end{itemize}
Since all four conditions are satisfied, the general convergence theorem for overfitted mixture models applies directly, guaranteeing the contraction rate $\mathcal{O}((\log N / N)^{1/2})$ and strong consistency for the number of clusters $k \rightarrow k_0$.
\end{proof}

\subsection{Asymptotic Properties of Cluster Partitions}
To further demonstrate the advantage of MFM over the traditional Dirichlet Process (DP) in the context of basket trials, we examine the asymptotic behavior of the number of active clusters, denoted as $T_N$, as $N \rightarrow \infty$.

\begin{prop}
  Under the DP prior with concentration parameter $\alpha > 0$, the expected number of clusters grows logarithmically with the number of cohorts $N$, such that $E[T_N \mid \text{DP}] \sim \alpha \log N$. Conversely, under the MFM prior with a light-tailed prior $q(k)$ on the number of components, $T_N$ converges almost surely to the true number of components $k_0$.  
\end{prop}

\begin{proof}
Under the DP Polya urn scheme in \eqref{eq:crp}, the probability of opening a new cluster at step $i$ is $\frac{\alpha}{\alpha + i - 1}$. Summing these probabilities yields the harmonic number sequence, proving $E[T_N] \approx \alpha \log N$, which diverges as $N \rightarrow \infty$. Under the MFM framework \eqref{eq:mcrp}, the probability of opening a new cluster depends on the ratio $\frac{V_i(t+1)}{V_i(t)}$. As shown by \cite{miller2018mixture}, for a Poisson prior on $k$, this ratio decays rapidly as $i$ increases, asymptotically forcing the creation of new clusters to zero once $T_N$ hits the true number of components $k_0$, ensuring almost sure convergence $T_N \xrightarrow{a.s.} k_0$.
\end{proof}

\subsection{Proposed two-step procedure}

We now propose a two-step procedure to find the balance between pooled and stratified analysis in basket trial design. In the first step, we treat it as a clustering problem by grouping cohorts into clusters that share the similar treatment effect using the model proposed in (\ref{eq:basket_MFM}). In the second step, we further address heterogeneity for cohorts within each cluster by fitting the Bayesian Hierarchical model by \cite{berry2013bayesian} in each cluster. For cluster $s$, the model and prior can be expressed as:
\begin{align}\label{eq:basket_BHM}
& \theta_j = \text{log}(\frac{p_j}{1-p_j}) - \text{log}(\frac{p_T}{1-p_T}) 
\\ \nonumber
& \theta_j \sim \mbox{N}(\mu,\sigma^2), \quad j =
1,
\ldots, n_s,  \\ \nonumber
& \mu \sim \mbox{N}(0,2^2), \quad \sigma^2 \sim \mbox{HN}(0,1), \\ \nonumber
& r_j \mid n_j, p_j \stackrel{\text{ind}}
\sim
\mbox{Binomial}(n_j, p_j), \quad j =
1,
\ldots, n_s, 
\end{align}
where $n_s$ denotes the number of cohorts in cluster $s$.

\section{Bayesian Inference}\label{sec:bayesian}

\subsection{MCMC Algorithm}\label{sec:MCMC}
For step one of our proposed methods, we design an efficient collapsed Gibbs sampler algorithm. By integrating out the mixture weights $\bm{\pi}$ analytically, we avoid the need for complex reversible jump MCMC algorithms, making it much easier to obtain the posterior samples for the clustering labels.

The Gibbs sampler for step one is presented in Algorithm~\ref{algorithm1}.

\begin{algorithm}[tbp]
  \caption{Collapsed sampler for MFM-BD}
  \label{algorithm1}
  \begin{algorithmic}[1]
  \Procedure{c-MFM-BD} {}
  \\ Initialize $\bm{z} = (z_1, \ldots, z_N)$ and $\bm{P} = (P_{s})$.
  \For{each iter $=1$ to $\mbox{M}$ }
  \\ Update $\bm{P} = (P_{s})$ conditional on $z$ in a closed form as:
  \begin{align*} 
 P_{s} \mid \bm{r},\bm{n},z & \sim \mbox{Beta}(\alpha + r_{[s]}, \beta + N_s - r_{[s]})
  \end{align*}   
  
  Where ${r}_{[s]}=\sum_{z_i=s}r_i$ and ${N}_{s}=\sum_{z_i=s}n_i$, $r=1,\ldots,k$. Here $k$ is the number of clusters formed by current $z$.
  
  \\ Update $\bm{z} = (z_1, \ldots, z_N)$ conditional on $\bm{P} = (P_{s})$, for each $i$ in $(1,\ldots,N)$, we can get a closed form expression for $P(z_i = c \mid z_{-i}, \bm{r},\bm{n}, \bm{P})$:
  \[ \propto \left\{
  \begin{array}{ll}
        [\abs{c} + \gamma] \mbox{ } \text{dbinom}(r_i,n_i,P_{z_i}) & \text{at an existing table $c$} \\
        \frac{V_N(\abs{\mathcal{C}_{-i}}  +1)}{V_N(\abs{\mathcal{C}_{-i}})}
        \gamma  m(\mathscr{S}_i) & \text{if $c$ is a new table} \\
  \end{array} 
  \right. \]
  where $\mathcal{C}_{-i}$ denotes the partition obtained by removing $z_i$ and 
  \begin{align*}
  m(\mathscr{S}_i) =  \binom{n_i}{r_i} \frac{B(r_i+\alpha,n_i-r_i+\beta)}{B(\alpha,\beta)}
  \end{align*}
  Where $B(\cdot)$ is the beta function.
  
  \EndFor
  \EndProcedure
  \end{algorithmic}
\end{algorithm}

To rigorously justify the transition probabilities utilized in the updates for $\bm{z}$, we establish the following proposition detailing the derivation of the collapsed sampler.

\begin{prop} {Validity of the Collapsed MFM Transition Probabilities}
The full conditional distribution of the cluster assignment $z_i$ given the remaining assignments $\bm{z}_{-i}$, the data $\bm{r}$, and the hyperparameters leaves the target marginal posterior $p(\bm{z} \mid \bm{r}, \bm{n})$ invariant.
\end{prop}

\begin{proof}
  Let $\bm{z}_{-i}$ denote the cluster assignments for all cohorts except cohort $i$. By Bayes' theorem, the full conditional probability is proportional to the product of the prior predictive distribution of $z_i$ and the marginal likelihood of the data $r_i$:
\begin{align*}
P(z_i = c \mid \bm{z}_{-i}, \bm{r}, \bm{n}) \propto P(z_i = c \mid \bm{z}_{-i}) \times P(r_i \mid z_i = c, \bm{r}_{-i}, \bm{n})
\end{align*}
From the P\'{o}lya urn scheme of the MFM in Equation \eqref{eq:mcrp}, the prior predictive probability $P(z_i = c \mid \bm{z}_{-i})$ is proportional to $\abs{c} + \gamma$ for an existing cluster $c$, and proportional to $\gamma \frac{V_N(\abs{\mathcal{C}_{-i}} + 1)}{V_N(\abs{\mathcal{C}_{-i}})}$ for a new cluster.   
\end{proof} 

For the likelihood term, if $c$ is an existing cluster with instantiated parameter $P_c$, the likelihood is simply the Binomial mass function evaluated at $P_c$. If $c$ is a new cluster, the parameter $P_c$ is not yet instantiated. Therefore, we must marginalize out the Beta prior to obtain the prior predictive distribution for the new observation. This yields a Beta-Binomial distribution:
\begin{align*}
m(\mathscr{S}_i) = \int_{0}^{1} \text{Binomial}(r_i \mid n_i, P) \text{Beta}(P \mid \alpha, \beta) dP = \binom{n_i}{r_i} \frac{B(r_i+\alpha,n_i-r_i+\beta)}{B(\alpha,\beta)}
\end{align*}
Multiplying the MFM prior predictive probabilities by these corresponding likelihood terms yields the exact conditional probabilities utilized in Algorithm \ref{algorithm1}. This confirms that integrating out the Dirichlet mixing weights $\bm{\pi}$ preserves the exact stationary distribution of the Markov chain. $\blacksquare$

For step two, we do not derive a custom MCMC algorithm since the full conditional distributions for the hierarchical logit-normal model are not in a conjugate closed form. Instead, the \texttt{gMAP} function in the \texttt{RBesT} package in R is utilized to handle Berry's BHM efficiently via Stan's No-U-Turn Sampler (NUTS).  

\subsection{Post MCMC inference}\label{sec:postmcmc}
Another important task for our proposed method is the inference of MCMC results. We first discuss the inference on the cluster assignment parameter $\bm{z}$. The standard posterior mean or median of $\bm{z}$ is not suitable due to the well-known ``label switching" problem in Bayesian mixture models, where the arbitrary indexing of clusters makes direct averaging of labels meaningless.

Dahl's method \citep{dahl2006model} provides a robust remedy for the posterior inference of the clustering configurations~$\bm{z}$. It chooses the iteration in the posterior sample that optimizes a least-squares criterion as the final estimate for $\bm{z}$. We first define an $N \times N$ membership matrix for a given MCMC draw: 
\begin{align}\label{eq:membermat}
  B = (B(i,j)) =
  \big( 1(z_i = z_j)\big), \quad i,j\in \{1,2,\dots,N\},
\end{align}
where $1(\cdot)$ is the indicator function. 
The $(i,j)$-th element of the membership matrix is $1$ if the $i$-th and $j$-th observations belong to the same component; it is $0$ otherwise. For the MCMC samples, we calculate $B^{(l)}$ for the $l$-th iteration. Then, we calculate the element-wise mean of the membership matrices $\bar{B} = \frac{1}{L} \sum_{l=1}^{L} {B}^{(l)}$, which represents the pairwise posterior probability that two cohorts are co-clustered. 

Finally, we identify the most \emph{representative} posterior draw as the one whose membership matrix is closest to $\bar{B}$ with respect to the element-wise Euclidean distance:
\begin{align*}
\arg\min_{l \in \{1,\ldots,L\}} \sum_{i=1}^{N} \sum_{j=1}^{N} \left( B^{(l)}(i,j) - \bar{B}(i,j) \right)^{2}
\end{align*}
The posterior estimates of the cluster memberships $z_1,\ldots,z_N$ are directly extracted from this optimal draw $l$ identified by Dahl's method. This guarantees that the final clustering estimate is an actual, valid partition generated by the model.

Within each cluster identified by step one, the cohort-specific parameter $\theta_j$'s will be summarized by their posterior means. The Bayesian estimate for the cohort response rate $p_j$ will then be directly obtained by solving the inverse-logit equation defined in \eqref{eq:basket_BHM}.

\section{Simulation}\label{sec:simu}
In this section, we investigate the performance of our proposed method from a variety of measures.

\subsection{Simulation Settings and Evaluation Metrics}\label{sec:simu_setup}
In the simulation study, we have five different scenarios with two different sample sizes (20 or 30) in each cohort. The response rates for each scenario are shown in Table~\ref{simulation_design}.
\begin{table}[htp]
    \centering
    \begin{tabular}{cc}
    \midrule
    Scenario& Response Rate Design\\
    \hline
        Scenario 1 & $P_1=P_2=\ldots=P_{10}=0.4$  \\
         Scenario 2& $P_1=\ldots=P_5=0.2$, $P_6=\ldots=P_{10}=0.6$ \\
           Scenario 3 & $P_1=\ldots=P_5=0.2$, $P_6=\ldots=P_{10}=0.5$ \\
         Scenario 4& $P_1=P_2=P_3=0.1$, $P_4=P_5=P_{6}=0.4$, $P_7=\ldots=P_{10}=0.7$\\
           Scenario 5 & $P_1=P_2=\ldots=P_{10}=0.2$ \\
       \bottomrule
    \end{tabular}
    \caption{Simulation designs for different scenarios}
    \label{simulation_design}
\end{table}

In all the simulation scenarios considered below, we employed Algorithm 1 with $\gamma$ = 1 and $\alpha = \beta = 1$ to fit the MFM-BD model. We arbitrarily initialized our algorithm with 5 clusters and randomly allocated the cluster assignments. We experimented with various other choices and did not find any evidence of sensitivity to the initialization. Results from our method is based on 5000 MCMC iterations leaving out a burn-in of 2000 in step one and 8000 MCMC iterations leaving out a burn-in of 2000 in step two.

For Berry's BHM and EXNEX approach, the choice of hyper-parameters are from the default settings as stated in section \ref{existing}. Results from those two approaches are based on 10000 MCMC iterations leaving out a burn-in of 2000. 

For our proposed method, the estimated number of clusters $\hat{K}$ for each replicate is summarized from the MCMC iteration picked by Dahl's method which is introduced in Section~\ref{sec:postmcmc}. The performance of the posterior estimates of parameters for all methods were evaluated by the average absolute bias (AAB) and the mean square error (AMSE) in the following ways, take $\theta_i$ as an example:
\begin{equation*}
\begin{split}
\text{AAB} =\frac{1}{10} \sum_{i=1}^{10}\left| \frac{1}{500} \sum_{r=1}^{500}(\theta^{r}_i-\theta^{0}_i)\right|, \\
\text{MMSE} =\frac{1}{10} \sum_{i=1}^{10}  
\sqrt{\frac{1}{500}
	\sum_{r=1}^{500} (\theta^{r}_i-\theta^{0}_i)^2},    
\end{split}
\end{equation*}

where $\theta^{r}_i$ is the posterior mean of rth replicate for ith cohort and $\theta^{0}_i$ is the true value of ith cohort. 

\subsection{Simulation Results}
First, we present the cluster estimates of our proposed methods in Table~\ref{tab:group_estimate}. From the results shown in Table~\ref{tab:group_estimate}, we see that our proposed methods successfully recover the number of clusters within a reasonable range for all five different scenarios under two different sample size. In addition, the large sample size will cause better estimates of the number of clusters by comparing results between $n=20$ and $N=30$.  
\begin{table}[htp]
	\centering
	\caption{Clustering Performance for Simulation Studies}\label{tab:group_estimate}
	\begin{tabular}{cccc}
		\toprule
	 Scenario&n&  $\hat{K}$& S.D. of $\hat{K}$  \\
		\midrule 
		 Scenario 1 & $n=20$ &1.046& 0.210 \\
		&$n=30$&1.036&0.186 \\
		 Scenario 2 &$n=20$ &2.168& 0.417 \\
		&$n=30$&2.131&0.346 \\
         Scenario 3 &$n=20$ &1.945& 0.516 \\
		&$n=30$&2.105&0.376 \\
		 Scenario 4 &$n=20$ &2.661& 0.612 \\
		&$n=30$&2.929&0.559 \\
		 Scenario 5 &$n=20$ &1.026& 0.159 \\
		&$n=30$&1.020&0.140 \\

		\bottomrule
	\end{tabular}
\end{table} 

Furthermore, we compare our proposed methods with Berry's BHM and EXNEX under the criteria we proposed in Section~\ref{sec:simu_setup}. The simulation results are presented in Table~\ref{table_simu_results20} and Table~\ref{table_simu_results30}. All simulations results are based on 500 replicates. From the results shown in Table~\ref{table_simu_results20} and Table~\ref{table_simu_results30}, we see that our proposed methods have better estimation performance based on AAB when heterogeneity exits among different cohorts and the AMSE is comparable with benchmark methods.  
\begin{table}[htp]
\centering
\caption{Simulation Results ($n=20$) }
\begin{tabular}{ccccc}
\toprule
                            &     & Berry's BHM & EXNEX & MFM-BD \\
                            \midrule
{Scenario 1} & AAB  &  0.0015       &  0.0026       & 0.0025         \\
                            & AMSE & 0.0024        &  0.0056        & 0.0026         \\
{Scenario 2} & AAB  &   0.0430    &     0.0274     & 0.0131        \\
                            & AMSE &  0.0091       &  0.0095        &  0.0069        \\
{Scenario 3} & AAB  &   0.0485        &   0.0304     &   0.0266      \\
                            & AMSE &   0.0087      & 0.0093        &  0.0096\\
{Scenario 4} & AAB  &  0.0305      &   0.0256       &     0.0261      \\
                            & AMSE & 0.0086        &  0.0086        &      0.0102    \\
{Scenario 5} &AAB  &  0.0023       &   0.0064       &     0.0119     \\
                            & AMSE &0.0016         & 0.037         &   0.0016       \\
                            \bottomrule
\end{tabular}
\label{table_simu_results20}
\end{table}

\begin{table}[htp]
\centering
\caption{Simulation Results ($n=30$) }
\begin{tabular}{ccccc}
\toprule
                            &     & Berry's BHM & EXNEX & MFM-BD \\
                            \midrule
{Scenario 1} & AAB  &   0.0012       &     0.0019    & 0.0015         \\
                            & AMSE & 0.0016        &0.0036          &  0.0017        \\
{Scenario 2} & AAB  &   0.0303      &     0.0180     &   0.0056       \\
                            & AMSE &  0.0063       &  0.0064        &      0.0035    \\
{Scenario 3} & AAB  &  0.0349      &     0.0223     &    0.0107      \\
                            & AMSE &  0.0061       &0.0065          & 0.0053       \\
{Scenario 4} & AAB  &      0.0207    &      0.0179    &    0.0136     \\
                            & AMSE &  0.0059       & 0.0060         &  0.0061        \\
{Scenario 5} &AAB  &     0.0015    &     0.0039   &      0.0068     \\
                            & AMSE &0.0011         &  0.0024        &  0.0010        \\
                        \bottomrule
\end{tabular}
\label{table_simu_results30}
\end{table}

\begin{figure}
     \centering
     \begin{subfigure}{0.48\textwidth}
         \centering
         \includegraphics[width=\textwidth]{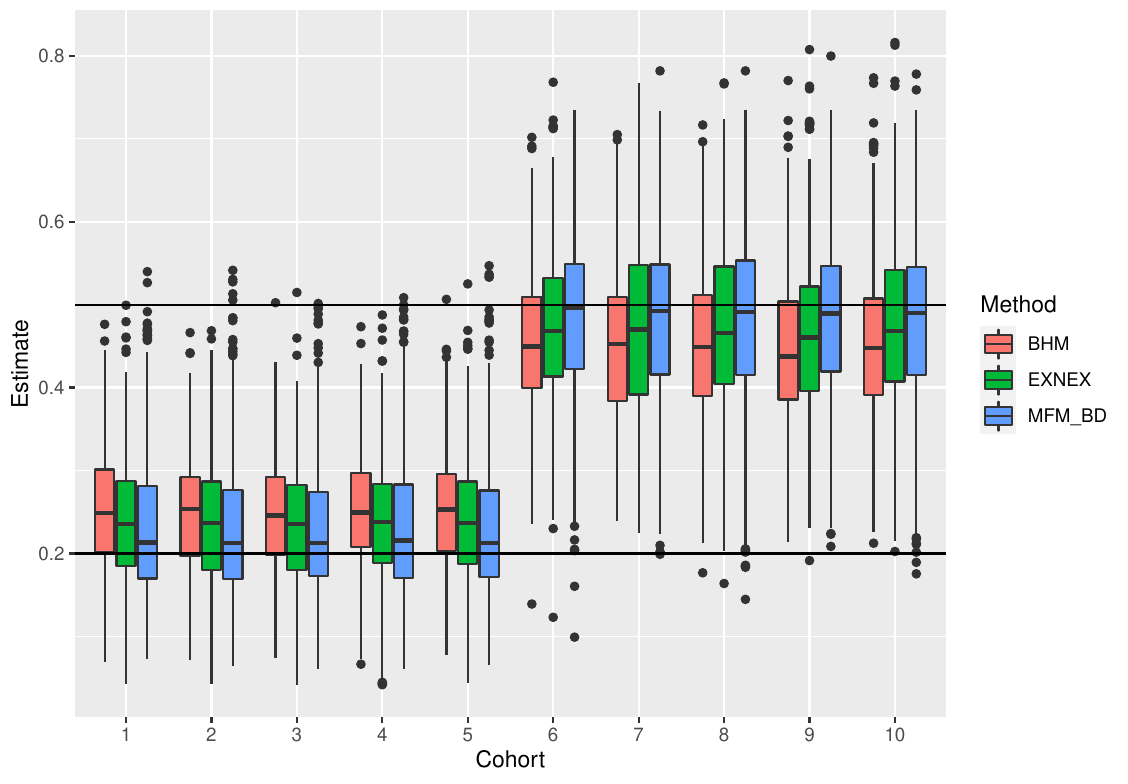}
         \caption{$n=20$}
         \label{fig:n20}
     \end{subfigure}
     \hfill
     \begin{subfigure}{0.48\textwidth}
         \centering
         \includegraphics[width=\textwidth]{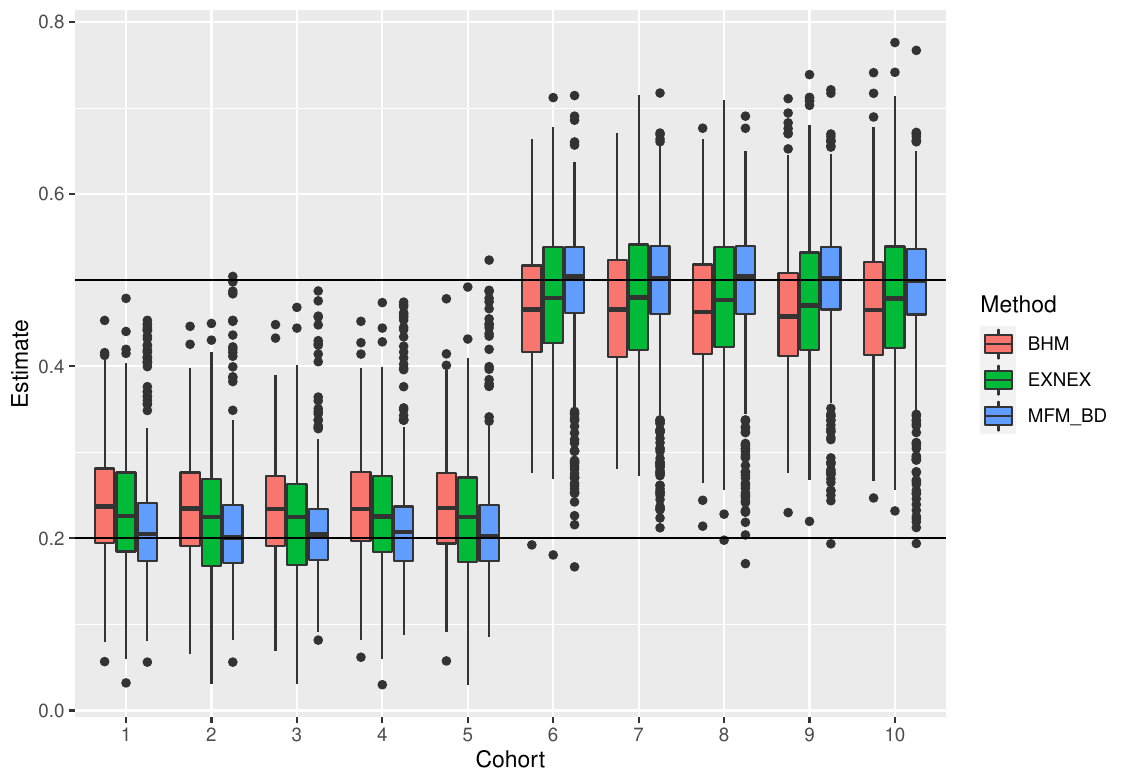}
         \caption{$n=30$}
         \label{fig:three sin x}
     \end{subfigure}
        \caption{Posterior Estimates Boxplots for Different Cohorts under Scenario 3}
        \label{fig:three graphs}
\end{figure}

In order to have a closer look of estimation performance, we have the boxplots of posterior estimates of different cohorts in Figure~\ref{fig:three graphs}. 
From the results shown in Figure~\ref{fig:three graphs}, we see that in each cohort our proposed method performs better than other two methods when there exit heterogeneities among different cohorts.
\section{Real Data Analysis}\label{sec:real_data}
In this section, we apply the proposed MFM-BD to the analysis of a famous basket trial \citep{hyman2015vemurafenib}. This trial is a histology-independent “basket” study of vemurafenib in BRAF V600 mutation–positive nonmelanoma cancers, where patients were enrolled in six prespecified cancer cohorts including anaplastic thyroid cancer (ATC), Erdheim-Chester disease or Langerhans’-cell histiocytosis (ECD/LCH), cholangiocarcinoma (CCA), colorectal cancer treated by vemurafenib (CRC-V), colorectal cancer treated by vemurafenib and cetuximab (CRCVC), and non small cell lung cancer(NSCLC). The primary end point was the
response rate. A total of 84 evaluable patients are included in the analysis. The sample size and observed responses for each cohort are presented in Table \ref{tab:real_data}. 

Results from our method (MFM-BD) is based on 5000 MCMC iterations leaving out a burn-in of 2000 in step one and 8000 MCMC iterations leaving out a burn-in of 2000 in step two;  initialized at a randomly generated configuration with 5 clusters. From Table \ref{tab:real_data}, 6 subtypes are grouped into two clusters. The first cluster includes ATC, ECD/LCH and NSCLC and the the second cluster includes CCA, CRC-V and CRC-VC. 

For comparison, we also conduct analysis based on Berry's BHM and EXNEX approach. The parameter settings for both methods are consistent from those used in simulation studies. From the results in Table \ref{tab:real_data}, we can see that when comparing to the other approaches, our proposed MFM-BD give the estimations more shrinkage within each cluster and more divergence between clusters.


\begin{table}
\caption{Response estimations Results}
\label{tab:real_data}
\centering
\begin{tabular}{llcclll}
\toprule
\multicolumn{2}{c}{\multirow{2}{*}{Subtype}}  & \multirow{2}{*}{$r/n$} & \multirow{2}{*}{$\%$} & \multicolumn{3}{c}{Posterior mean (CI) } \\ \cline{5-7}
& &             &                                         &MFM-BD    &         Berry's BHM     &  EXNEX    \\ \midrule
\multirow{3}{*}{\rotatebox[origin=c]{90}{\footnotesize{Cluster 1}}} &1.          ATC       &      2/7            &          28.6          &     31.8 (6.9,62.3)      &      26.7 (6.3,57.6)    &      27.5 (5.1,60.1)      \\
 &2.          ECD/LCH       &       6/14            &         42.9          &     41.6 (20.3,65.7)      &     38.3 (17.7,63.2)     &     40.7 (18.1,65.3)           \\
  &6.         NSCLC        &       8/19            &             42.1      &    41.2 (21.9,61.9)       &    38.6 (19.8,60.5)      &       40.5 (20.6,62.3)    \\
   \addlinespace[2ex]

\multirow{3}{*}{\rotatebox[origin=c]{90}{\footnotesize{Cluster 2}}} &3.     CCA          &          1/8           &          12.5         &   12.6 (1.7,35.7)        &    16.2 (2.1,39.9)      &       15.4 (1.5,42.5)       \\
&4.       CRC-V          &          1/26         &            3.8       &     6.6 (0.7,16.7)      &   7.5 (0.9,20.3)       &      5.9 (0.6,16.7)      \\
 &5.     CRC-VC            &         0/10          &          0.0         &     6.5 (0.1,20.4)      &     8.3 (0.2,26.5)     &       5.8 (0.1,23.3)      \\

                  \bottomrule
\end{tabular}
\end{table}
\section{Discussion}\label{sec:discussion}

In this article, we propose a novel two-step Bayesian procedure to evaluate the efficacy of a new targeted treatment (using a binary response endpoint) across multiple disease cohorts in a basket trial. Basket trials inherently face a critical statistical dilemma: complete pooling of cohorts risks a highly inflated Type I error if the treatment only works in a subset of indications, while strictly independent analyses suffer from low statistical power due to the typically small sample sizes in early-phase oncology. Our proposed framework navigates this trade-off effectively. It demonstrates excellent performance in both extensive simulation studies and real data applications by striking a balance between information borrowing and strict stratification. 

The primary innovation of this article lies in its data-driven approach to borrowing information. Rather than forcing a global exchangeability assumption across all cohorts, our method first clusters the cohorts into distinct groups without needing to pre-specify the number of groups, utilizing the Mixture of Finite Mixtures (MFM) model. We then borrow information strictly within each resulting group under a localized exchangeability assumption. Beyond the statistical advantages, the clustering of these cohorts provides highly actionable clinical directions for the future development of a new drug. For instance, if diverse tumor histologies consistently cluster together into a high-response group, it generates strong biological hypotheses regarding shared genomic profiles or overlapping mechanism-of-action pathways, guiding subsequent Phase III trial designs. Furthermore, unlike the traditional Dirichlet Process, our approach is theoretically proven to yield consistent detection of the true number of clusters, actively pruning the extraneous, clinically meaningless micro-clusters that often plague Bayesian non-parametric models.

While we highly recommend using Berry's BHM in the second step of our approach due to its computational efficiency and widespread acceptance in regulatory settings, the framework is highly modular. Other popular hierarchical approaches, such as the EXNEX (Exchangeable-Nonexchangeable) model, can seamlessly be implemented in the second step. Utilizing EXNEX within the identified clusters could serve as an additional safeguard, addressing any residual heterogeneity or extreme outlying responses that might exist even among broadly similar cohorts. 

Despite its robust performance, this approach does have certain limitations. The proposed MFM-based clustering is most appropriate for basket trials where the number of cohorts is relatively large (e.g., 6 cohorts or more). In highly restricted trials with only three or four cohorts, the clustering algorithm may lack the necessary data diversity to form stable, non-trivial partitions, often defaulting to either complete pooling or complete independence. Additionally, the reliance on Dahl's method for post-MCMC inference requires storing large membership matrices, which can be computationally intensive for massive scale simulations, though it remains highly tractable for standard trial sizes.

There are several promising directions for the further extension of this work. First, while our current model focuses on binary endpoints (e.g., Objective Response Rate), many later-stage trials rely on time-to-event endpoints. Extending this MFM framework to analyze Progression-Free Survival (PFS) and Overall Survival (OS) \citep{xu2019nonparametric} would require integrating survival likelihoods (such as Weibull or piecewise exponential models) and accounting for right-censoring within the clustering step. Second, we aim to add another layer of hierarchical heterogeneity modeling so that this method can be deployed in complex platform trials. Unlike basket trials, platform trials evaluate multiple treatment options simultaneously against multiple cancer types, requiring a multi-dimensional clustering approach to capture treatment-by-disease interactions. Finally, incorporating patient-level biomarker covariates directly into the MFM prior—allowing baseline genetic signatures to influence the probability of two cohorts clustering together—could further refine the precision of the dynamic borrowing process.

\bibliographystyle{chicago}
\bibliography{model}
\end{document}